\documentstyle[preprint,prl,aps]{revtex}
\begin{document}

\title{Comment on "Magnetic Relaxations of Antiferromagnetic
Nanoparticles in Magnetic Fields"}
\author{E. del Barco, M. Duran, J. M. Hernandez, and J. Tejada}
\address{Departamento de F\'\i sica Fundamental, Universidad de
Barcelona. Diagonal 647, Barcelona, 08028, Spain}
\maketitle

In a recent Letter \cite{Mamiya}, Mamiya et al. presented a new
and elegant experimental procedure to control the initial
Field-Cooled Magnetization, $M_{FC}$, of the magnetic relaxations
of ferritin under study. They use a step-by-step cooling from $T$=
35 K to the relaxation experimental temperature, $T_0$= 8-9 K,
while a magnetic field, $H=H_0T/T_0$, is applied at each step to
eliminate time- and cooling rate-dependences of $M_{FC}$. From the
temperature behavior of the time magnetic relaxation curves,
recorded at temperatures between 8 K and 9 K, they concluded that
relaxation in ferritin is essentially dominated by classical
activated process. Moreover, they analyzed the field dependence of
the magnetic viscosity extracted from the slopes of the relaxation
curves recorded at different final fields, $H_f$= 0,-$H_0$. They
found that magnetic viscosity at $T$= 8 K was independent of
field. The authors finally concluded that {\it magnetic
relaxations of ferritin in magnetic fields are dominated by
classical superparamagnetic fluctuations in the temperature regime
where thermally assisted Resonant Spin Tunneling (RST) has been
discussed in recent years}.

Let us first to point out that we agree with the interpretation
the authors do of their results in the temperature range 8-9K in
terms of thermal activated relaxation. However, it is of great
importance to clarify that {\it there is not any paper} (from
those cited by Mamiya et al.) {\it in which activated quantum
tunneling has been used to explain relaxation phenomena of
ferritin in this high temperature range}, $T\sim$ 8-9 K. From
magnetic viscosity measurements, Tejada et al. \cite{Tejada}
reported resonant quantum tunneling at zero field at temperatures
around 4-5 K. These results were confirmed by Luis et al.
\cite{Luis} by a.c. susceptibility measurements at low
temperature. It is clear that a decrease of the viscosity with
small fields around the blocking temperature, $T_B$ = 8 K, is due
to the large variation of the number of unblocked magnetic
moments. The same phenomena was observed at lower temperatures
with higher fields \cite{Tejada}, where $S(H)$ shows a maximum at
a given field which decreases when the temperature increases.
Recently, new experimental results on magnetic relaxation of
ferritin confirm both pure quantum tunneling behavior between 0.1
K and $\sim$2 K and thermally activated quantum tunneling up to
$\sim$5 K \cite{Duran}, through magnetic viscosity measurements in
the absence of magnetic fields in the relaxation processes.

We have also used both the Field-Cooling method and data analysis
of Mamiya et al. \cite{Mamiya} to study time relaxation in
ferritin at $T$= 5 K, where RST was reported \cite{Tejada}. We
have used {\it FLUKA} commercial diluted ferritin. We have
measured time magnetic relaxation from different initial fields,
$H_0$, to two final fields, $H_f$= 0, -$H_0$, from $H_0$= 5 Oe up
to $H_0$= 800 Oe. In figure 1 we show the field dependence of the
ratio between the slopes of the relaxation curves for $H_f$=
-$H_0$ and $H_f$= 0, normalized with 1/2 to take into account the
difference between the equilibrium magnetizations. Figure 1 is
divided in two zones: For $\frac{S_{H_f=-H_0}}{ S_{H_f=0}}>1$
(gray zone), the system relaxes faster with field than without
field, indicating that the relaxation is governed by thermal
processes; For $\frac{S_{H_f=-H_0}}{S_{H_f=0}}<1$ (white zone),
the relaxation is faster at zero field, in agreement with RST
interpretation. As one can observe in figure 1, at $T$= 5 K, zero
field relaxation in ferritin is bigger than field relaxation for
fields below 500 Oe. The biggest difference ($\sim$80$\%$) is
observed at $H$= 5 Oe, where the ratio between viscosities is
minimum, in good agreement with the zero field resonant width,
$\Delta H_{eff}\sim$5 Oe, found through Landau-Zener relaxation
experiments by Duran et al. \cite{Duran}.

\pagebreak

{\bf FIGURE CAPTIONS}\\

{\bf Figure 1:} Ratio between non-zero-field and zero-field
viscosities as a function of the magnetic field, $H_0$, recorded at $T$= 5 K in ferritin.\\

\end{document}